\documentclass[conference]{IEEEtran}
\usepackage{amsfonts}
\usepackage{cite}
\usepackage{amsmath}
\usepackage{amssymb}
\usepackage[dvips,final]{epsfig}
\usepackage{color}
\usepackage{graphicx}
\usepackage{stfloats}
\usepackage{algorithm}
\usepackage{algorithmic}

\makeatletter
\let\thanks\@IEEESAVECMDthanks
\long
\def\@makefntext#1{\parindent 1em\indent\hbox{\@makefnmark}#1}
\renewcommand\footnoterule{  \kern-3\p@
  \hrule\@width.4\columnwidth
  \kern2.6\p@}
\makeatother 
\begin{document}
\title{Relay Assisted Cooperative OSTBC Communication with SNR Imbalance and Channel Estimation Errors
\thanks{This work was conducted while the first author was employed as a summer intern with the CTO Office, InterDigital.} }
\vspace{-1.80in}
\author{\authorblockN{Bo Niu}\\
\authorblockA{Department of Electrical and Computer Engineering, \\New
Jersey Institute of Technology \\Newark, NJ 07102, USA
\\ bo.niu@njit.edu}\\ \and
\authorblockN{Mihaela C. Beluri, Zinan Lin, Prabhakar Chitrapu}\\
\authorblockA{InterDigital\\
King of Prussia, PA 19406, USA
\\ \{mihaela.beluri, zinan.lin, prabhakar.chitrapu\}@interdigital.com}}
\maketitle \vspace{-0.38in}
\begin{abstract}
In this paper, a two-hop relay assisted cooperative Orthogonal
Space-Time Block Codes (OSTBC) transmission scheme is considered for
the downlink communication of a cellular system, where the base
station (BS) and the relay station (RS) cooperate and transmit data
to the user equipment (UE) in a distributed fashion. We analyze the
impact of the SNR imbalance between the BS-UE and RS-UE links, as
well as the imperfect channel estimation at the UE receiver. The
performance is analyzed in the presence of Rayleigh flat fading and
our results show that the SNR imbalance does not impact the spatial
diversity order. On the other hand, channel estimation errors have a
larger impact on the system performance. Simulation results are then
provided to confirm the analysis.
\end{abstract}
\section{Introduction}
Cooperative diversity capitalizes on transmissions from antennas at
different nodes in order to provide spatial diversity, thus
enhancing the performance of a wireless network
\cite{Laneman:IT_04}. Utilizing distributed Orthogonal Space-Time
Block Codes \cite{Tarokh:IT_99_1}\cite{Alamouti:JSAC_98} has been
proven to be effective in realizing the cooperative diversity
\cite{Laneman:IT_03}\cite{Bolcskei:JSAC_04}. Various schemes have
been proposed for this
purpose\cite{Hassibi:TWC_06,Hassibi:IT_07,Rajan:IT_submitted,Cimini:Eurasip_08}.
Using relay to help the transmission is a popular choice. A virtual
antenna array is created with source nodes and relay nodes to
implement the cooperative transmission. The relay can be either
amplify-and-forward (AF) or decode-and-forward (DF). Detail analysis
of this two schemes can be found in
\cite{Laneman:IT_04}\cite{Laneman:IT_03}.

In our recent work \cite{Chitrapu:Tdoc_08}, we proposed a relay
assisted cooperative transmission scheme for the Long Term Evolution
(LTE)-Advanced system. It has been shown that this scheme, by
employing RS in the cellular network and working with a two-hop
cooperative diversity scheme, significantly improves the system
coverage and capacity. In the first hop of the transmission scheme,
both the RS and UE receive data from the BS. We assume that the BS
has a better channel condition to the RS than to the UE. Thus the RS
successfully decodes the data before the UE does. In the second hop,
the BS and RS cooperate and transmit data to the UE. The UE then
decodes the received data that is transmitted from both the BS and
RS. Details of the relay assisted transmission scheme can be found
in \cite{Chitrapu:Tdoc_08}. In this paper, we focus on the
cooperative transmission and its reception in the second hop.

The performance of the cooperative diversity transmission scheme
used in the second hop may be impacted by the impairments associated
with a practical receiver implementation as well as by the
distributed nature of the cooperation scheme. The impairments
analyzed in this paper are:    1) the SNR imbalance between the
BS-UE and RS-UE links, and 2) channel estimation errors at the UE
receiver. In practical deployment scenarios, the transmit powers of
the BS and the RS may be different. Additionally, the propagation
loss on the BS-UE link is likely different from the propagation loss
on the RS-UE link. The different transmit powers and propagation
losses for the two links, in conjunction with imperfect power
control, may lead to SNR imbalance between the BS-UE and RS-UE links
(with possibly a better SNR for the RS-UE link). This may result in
performance degradation as compared to the case of equal SNR per
each link. To evaluate the impact of channel estimation errors, an
MMSE estimator is used in the literature, and the channel estimation
error is assumed to be a zero mean complex Gaussian random variable
and independent of the channel \cite{Tarokh:IT_99_2}. The
performance of the OSTBC scheme with imperfect channel estimation
has been analyzed for the Alamouti case in \cite{Leung:EL_03}, but
to the best of our knowledge, not for the case of distributed
antennas with SNR imbalance and channel estimation errors.

\subsection*{Contributions and Relation to Previous Work}

In this work, we consider realistic scenarios for a relay assisted
wireless cellular system and shed light on issues that deserve
attention during practical network design. We study the performance
of a cooperative OSTBC diversity scheme with the consideration of
SNR imbalance and channel estimation errors, which are impairments
that are likely to occur in practice. The system setup is similar to
\cite{Chitrapu:Tdoc_08}, except that we introduce the OSTBC
transmission scheme in the second hop to achieve spatial diversity,
and improve the system performance. The previous result of
\cite{Leung:EL_03} is obtained as a special case of our work, for no
SNR imbalance. The detailed contributions reported in this paper can
be summarized as follows:

$\bullet$ Derivation of  the theoretical probability of error
performance (BER vs. SNR) for the cooperative $2\times 1$ OSTBC
(Alamouti) scheme with SNR imbalance between the RS-UE and the BS-UE
link.

$\bullet$ Proof that the SNR imbalance between the RS-UE and the
BS-UE link does not affect the spatial diversity of the cooperative
system.

$\bullet$ Simulation results that show the channel estimation
errors, compared to SNR imbalance, have a major effect on the
cooperative OSTBC system performance.

The rest of the paper is organized as follows. In the following
section, the relay assisted cooperative OSTBC communication system
is introduced. We especially focus on the second hop transmission of
the scheme where the BS cooperates with the RS and transmits data to
the UE. In Section III, we analyze the system performance
degradation due to SNR imbalance between the BS-UE and RS-UE links.
We derive the theoretical closed-form probability of error
expression for the cooperative Alamouti scheme with SNR imbalance,
and prove that the SNR imbalance does not affect the spatial
diversity of the scheme. After that, we investigate the effect of
channel estimation errors on the system performance in Section IV.
Simulation results are then provided in Section V to confirm our
analytical results. Concluding remarks are offered in Section VI.

\emph{Notation:} Throughout the paper, we denote matrices and
vectors with bold face type, using capital letters for matrices and
lower case letters for vectors. For any matrix $\mathbf{A}$, the
superscript $^{*}$ denotes complex conjugate transpose.

\section{System Model}
In this section, the system model for the relay assisted cooperative
OSTBC communication system is introduced. We consider the downlink
transmission in a cellular system, where a BS transmits data to UEs.
We assume that there is also a RS in the system in order to assist
the transmission. As we mentioned before we focus on the second hop
transmission of \cite{Chitrapu:Tdoc_08} when both the BS and RS have
a copy of the transmitted signal (for example, Decode-and-Forward
transmission is performed in the first hop and the RS has
successfully decoded the data transmitted from BS). The antennas at
the BS and RS create a virtual multiple antenna transmitter, encode
the data with OSTBC and transmit the coded data cooperatively
utilizing all their antennas. For the ease of the theoretical
analysis in the following sections, we investigate the cooperative
Alamouti scheme for our relay assisted communication. We assume the
BS and RS each have one transmit antenna and the UE has one receive
antenna. The case of cooperative OSTBC communication with $M_B, M_R$
transmit antennas at the BS and relay, respectively, and $N$ receive
antennas at the UE can be easily
 generalized.

With the system model introduced above, the received signal vector
$\textbf{y}=\begin{bmatrix} y_{t0} & y_{t1}
\end{bmatrix}$ at the UE can be represented as
\begin{eqnarray}
\textbf{y}=\sqrt{P_B}h_B\textbf{s}_B+\sqrt{P_R}h_R\textbf{s}_R+\textbf{z},
\label{eq:chn}
\end{eqnarray}
where $h_B$ and $h_R$ represent the Rayleigh fading channel gains
for the BS-UE and RS-UE links, respectively. Both of them have
i.i.d, complex Gaussian distributions with zero mean and unit
variance $\sigma_n^2$. $\textbf{s}_B$ and $\textbf{s}_R$ are the
$1\times 2$ transmitted signal vectors from the BS and RS,
respectively. They form the Alamouti code as
\begin{equation*}\textbf{S}=
\begin{bmatrix}
\textbf{s}_B \\
\textbf{s}_R
\end{bmatrix}=\begin{bmatrix}
s_0 & -s_1^*\\
s_1 & s_0^*
\end{bmatrix},
\end{equation*}
where $s_0,s_1$ are transmitted symbols drawn from an equuiprobable
source. $\textbf{z}$ is the received noise vector at the UE. Each
entry in $\textbf{z}$ is a complex Gaussian random variable with
zero mean and unit variance; $P_B$ and $P_R$ are the received powers
at the UE due to the transmission from the BS and RS, respectively.
The total power is $P=P_B+P_R$.

In a practical system, $P_B$ and $P_R$ are generally not the same,
which leads to the SNR imbalance in a relay assisted cooperative
system. The received SNR for the BS-UE link and RS-UE link are
available at the UE since it is able to perform channel estimation
of the BS-UE and RS-UE links separately based on different reference
symbols. The received reference pilot symbols from the BS and RS are
then scaled differently according to the transmitted power and the
path loss. When considering the SNR imbalance, we define as $r$ the
SNR ratio between the BS-UE and RS-UE links. Thus \eqref{eq:chn} can
be rewritten as
\begin{eqnarray}
\textbf{y}=\sqrt{P}\begin{bmatrix} \sqrt{\frac{1}{1+r}}h_B &
\sqrt{\frac{r}{1+r}}h_R
\end{bmatrix}\textbf{S}+\textbf{z}.\label{eq:chn1}
\end{eqnarray}

Upon receiving $\textbf{y}$, the UE performs the revised Alamouti
decoding, which considers the SNR imbalance $r$ between RS-UE and
BS-UE links, by using the estimated channel state information
$\hat{h}_B$ and $\hat{h}_R$. The decoded symbols are
\begin{equation}
\begin{bmatrix}
\tilde{s}_0 \\
\tilde{s}_1
\end{bmatrix}=\begin{bmatrix}
\sqrt{\frac{1}{1+r}}\hat{h}_B^* & \sqrt{\frac{r}{1+r}}\hat{h}_R\\
\sqrt{\frac{r}{1+r}}\hat{h}_R^* & -\sqrt{\frac{1}{1+r}}\hat{h}_B
\end{bmatrix}\begin{bmatrix}
y_{t0} \\
y_{t1}^*
\end{bmatrix}.\label{eq:decoded}
\end{equation}
In practice, there are errors during the channel estimation process.
Following \cite{Tarokh:IT_99_2}, we assume
\begin{eqnarray}
\hat{h}_B=h_B+n_B,\label{eq:h_B_hat}
\end{eqnarray}
and
\begin{eqnarray}
\hat{h}_R=h_R+n_R,\label{eq:h_R_hat}
\end{eqnarray}where $n_B$ and $n_R$ are the estimation errors, assumed to be independent zero mean complex Gaussian random
variables with variance $\beta$.

In the following sections, we investigate the effect of SNR
imbalance and channel estimation errors to the performance of the
relay assisted cooperative diversity scheme.

\section{Cooperative Alamouti diversity scheme with SNR Imbalance}
In this section, we analyze the impact of the SNR imbalance between
the BS-UE and RS-UE links. We assume that there are no channel
estimation errors at the UE, i.e., $\hat{h}_B=h_B$ and
$\hat{h}_R=h_R$.

\subsection{Performance Analysis for Cooperative Alamouti diversity scheme with SNR Imbalance}
Considering the SNR imbalance, the decoded symbol vector
\eqref{eq:decoded} can be rewritten as
\begin{eqnarray}
\begin{bmatrix}
\tilde{s}_0 \\
\tilde{s}_1
\end{bmatrix}
&=&\sqrt{P}\begin{bmatrix}
\left(\frac{1}{1+r}|h_B|^2+\frac{r}{1+r}|h_R|^2\right)s_0 \\
\left(\frac{r}{1+r}|h_R|^2+\frac{1}{1+r}|h_B|^2\right)s_1
\end{bmatrix}\nonumber\\
&+&\begin{bmatrix}
\sqrt{\frac{1}{1+r}}h_B^*z_0+\sqrt{\frac{r}{1+r}}h_Rz_1^*\\
\sqrt{\frac{r}{1+r}}h_R^*z_0-\sqrt{\frac{1}{1+r}}h_Bz_1^*
\end{bmatrix}.\label{eq:SNR_imb_decoded}
\end{eqnarray}
It can be readily shown that $\sqrt{\frac{1}{1+r}}h_B^*z_0$,
$\sqrt{\frac{r}{1+r}}h_Rz_1^*$ and $\sqrt{\frac{r}{1+r}}h_R^*z_0$,
$\sqrt{\frac{1}{1+r}}h_Bz_1^*$ are independent, zero mean Gaussian
random variables. Thus $\tilde{s}_0$ is the sum of
$\left(\frac{1}{1+r}|h_B|^2+\frac{r}{1+r}|h_R|^2\right)s_0$ and an
independent, zero mean Gaussian random variable with variance
$\left(\frac{1}{1+r}|h_B|^2+\frac{r}{1+r}|h_R|^2\right)$. Similarly,
$\tilde{s}_1$ is the sum of
$\left(\frac{r}{1+r}|h_R|^2+\frac{1}{1+r}|h_B|^2\right)s_1$ and an
independent zero mean Gaussian random variable with variance
$\left(\frac{r}{1+r}|h_R|^2+\frac{1}{1+r}|h_B|^2\right)$. Since
$s_0$ and $s_1$ are drawn from an equiprobable source, we have the
following proposition.

\textbf{Proposition 1} \emph{The probability of error performance of
the cooperative $2\times 1$ Alamouti scheme with SNR imbalance $r$
between the $RS-UE$ and the $BS-UE$ links is given by
\begin{eqnarray}
Pe&=&\frac{1}{2}\left(1-\frac{1}{\sqrt{1+\frac{2(1+r)}{a^2r \gamma}}}\right)\left(1-\frac{1}{\sqrt{1+\frac{2(1+r)}{a^2 \gamma}}}\right)\nonumber\\
&&\left(1+\frac{1}{\sqrt{1+\frac{2(1+r)}{a^2r
\gamma}}+\sqrt{1+\frac{2(1+r)}{a^2 \gamma}}}\right),\label{eq:Pe}
\end{eqnarray}
where $a$ is a constant that depends on the specific modulation mode
(e.g. for BPSK, $a=\sqrt{2}$ and for QPSK, $a=1$ ), and
$\gamma=P/\sigma_n^2$.}

\emph{Proof:} See Appendix A.
\subsection{Diversity of Cooperative Alamouti Scheme with SNR Imbalance}
We now look at the spatial diversity of the relay assisted
cooperative Alamouti scheme and investigate the effect of SNR
imbalance on the diversity order.

\textbf{Proposition 2} \emph{The diversity order of the relay
assisted cooperative Alamouti scheme with SNR imbalance between the
BS-UE and RS-UE links is 2, which is the same as the regular
Alamouti scheme. Thus the SNR imbalance does not affect the
diversity order.}

\emph{Proof:} We start from the definition of diversity order in
\cite{Zheng:IT_03},
\begin{equation}
d=-\lim_{\gamma\rightarrow \infty }\frac{\log Pe(\gamma)}{\log
\gamma}, \label{e:d}
\end{equation}
and the error of probability in \eqref{eq:Pe}, the diversity order
of the relay assisted cooperative Alamouti scheme with SNR imbalance
can be written as,
\begin{eqnarray}
d_s\!\!&=&\!\!-\!\!\lim_{\gamma\rightarrow \infty}\!\!
\frac{\log\!\!\left(\!1\!\!-\!\!\frac{1}{\sqrt{1+\frac{2(1+r)}{a^2r\gamma}}}\right)}{\log
\gamma}+\frac{\log\!\!\left(\!\!1\!\!-\!\!\frac{1}{\sqrt{1+\frac{2(1+r)}{a^2
\gamma}}}\right)}{\log
\gamma}\nonumber\\
&&+\frac{\log\left(1+\frac{1}{\sqrt{1+\frac{2(1+r)}{a^2r
\gamma}}+\sqrt{1+\frac{2(1+r)}{a^2\gamma}}}\right)}{\log
\gamma}\label{eq:diversity_mid}
\end{eqnarray}
At high SNR, the Taylor series expansion yields \cite{Tse:book_05}
\begin{eqnarray}
\sqrt{\frac{B}{A+B}}=1-\frac{A}{2B}+\mathcal{O}\left(\frac{1}{B^2}\right)\label{eq:taylor}
\end{eqnarray}
Introducing \eqref{eq:taylor} in \eqref{eq:diversity_mid}, it can be
easily shown that $d_s=2$ \hfill$\blacksquare$
\section{Cooperative Alamouti diversity scheme with SNR Imbalance and Channel Estimation Errors}
In this section, we look at the effect of both SNR imbalance and
channel estimation errors on the relay assisted cooperative Alamouti
scheme. To evaluate the impact of channel estimation errors, an MMSE
estimator is used in the literature, and the channel estimation
errors are assumed to be zero-mean complex Gaussian random variables
and independent of the channel \cite{Tarokh:IT_99_2}.

From \eqref{eq:h_B_hat} and \eqref{eq:h_R_hat}, $\hat{h}_B$ and
$\hat{h}_R$ are random variables with zero mean and variance
$\sigma_B^2=\sigma_R^2=1+\beta$, and $h_B$ and $\hat{h}_B$, as well
as $h_R$ and $\hat{h}_R$, are joint complex Gaussian distributed
with normalized correlation coefficient $\frac{1}{\sqrt{1+\beta}}$.
$h_B$ and $h_R$ can then be represented as
\begin{eqnarray}
h_B=\rho \hat{h}_B+w_B
\end{eqnarray}
and
\begin{eqnarray}
h_R=\rho \hat{h}_R+w_R
\end{eqnarray}
where $\rho=\frac{1}{1+\beta}$; $w_B$ and $w_R$ are complex Gaussian
random variables with zero means and variance
$\sigma_d^2=\frac{\beta}{1+\beta} $\cite{Leung:EL_03}. The decoded
symbol vector can be rewritten as \eqref{eq:decoded}.

Note that the previous result of \cite{Leung:EL_03} is obtained as a
special case when $r=1$.  In general it is difficult to find a
closed-form equation for the probability of error for the
cooperative Alamouti diversity scheme with SNR imbalance and channel
estimation errors. We show the effect of both by simulation in
Section V.
\section{Simulation Results}
Simulation results are provided to show how SNR imbalance and
channel estimation errors impact the cooperative OSTBC transmission
in i.i.d Rayleigh fading channels.

\begin{figure}%
\centering {\includegraphics[width=3.0in]{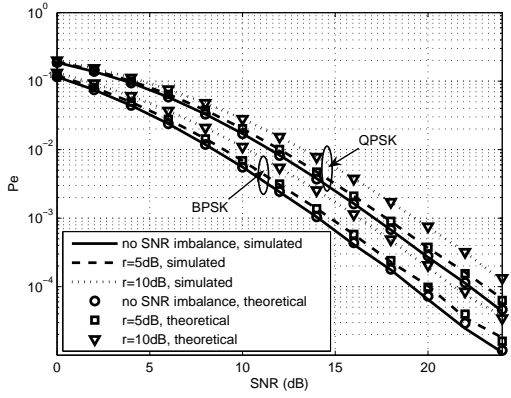}
\caption{BER vs. SNR for the cooperative Alamouti scheme for
different SNR imbalances $r=0,5$ dB and modulations BPSK and QPSK.}%
\label{fig:SNR_imbalance}}
\end{figure}

Fig.\ref{fig:SNR_imbalance} shows the theoretical and simulated BER
versus SNR performance of the cooperative OSTBC (Alamouti) scheme
when there is one receive antenna, with different SNR imbalances
$r=5, 10$ dB and different modulation modes (BPSK and QPSK), for the
case of no channel estimation errors. A performance curve for no SNR
imbalance ($r=0$ dB) case is also shown for reference. By allocating
different powers of $\frac{1}{1+r}\gamma$ and $\frac{r}{1+r}\gamma$
to each symbol at the BS and RS respectively, the results in Fig.
\ref{fig:SNR_imbalance} show that the performance degradation is
increasing with SNR imbalance. For an SNR imbalance of 10 dB, the
performance loss is measured around 2 dB for a probability of error
of $10^{-2}$. More generally, we see from the figure that the curves
drawn from the theoretical result \eqref{eq:Pe} perfectly match the
simulation curves, showing that the difference between the
probabilities of error performance with different SNR imbalance
ratio $r$ can be well quantified through our analysis of
\eqref{eq:Pe}. With regard to the spatial diversity, it is clear in
the figure at high SNR region that the cooperative Alamouti scheme
achieves a diversity of $2$, which is the same as the regular
Alamouti scheme.

\begin{figure}%
\centering {\includegraphics[width=3.0in]{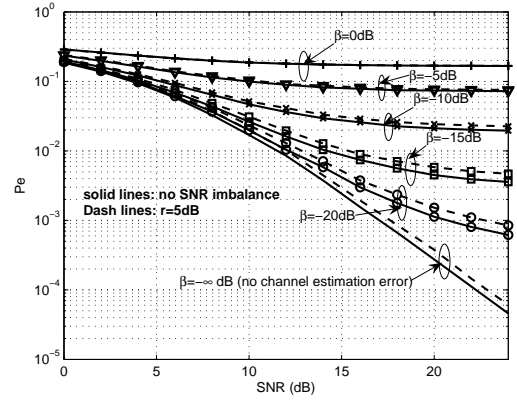}
\caption{BER vs. SNR for the cooperative Alamouti scheme for different SNR imbalances $r=0,5$ dB and channel estimation errors, QPSK modulation.}%
\label{fig:SNR_imbalance_che}}%
\end{figure}

\begin{figure}%
\centering {\includegraphics[width=3.0in]{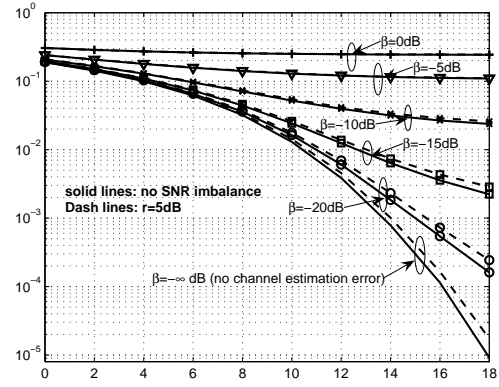}
\caption{BER vs. SNR for the cooperative OSTBC scheme ($M_B=M_R=N=2$) for different SNR imbalances $r=0,5$ dB and channel estimation errors, 16QAM modulation.}%
\label{fig:SNR_imbalance_che_4x2}}%
\end{figure}

Fig. \ref{fig:SNR_imbalance_che} compares the average BER versus SNR
for the cooperative Alamouti scheme employing QPSK modulation for
different levels of SNR imbalances and Gaussian channel estimation
errors. Performance of the cooperative OSTBC system with different
system setup (2 transmitters at both the BS and RS and 2 receive
antennas at the UE) is shown in Fig.
\ref{fig:SNR_imbalance_che_4x2}. It is seen that channel estimation
errors drastically affect the performance of the system, especially
at high SNR region, showing a more important role than the SNR
imbalance. Thus the channel estimation deserves more attention when
designing the system.

%

\section{Concluding Remarks}
In this paper, we studied the cooperative OSTBC communication in the
second hop of a two-hop decode-and-forward relay assisted cellular
network. We investigated the effect of SNR imbalance and channel
estimation error on the system performance. A closed-form expression
for the probability of error performance is derived to analyze the
performance degradation due to the SNR imbalance. It is also proved
that the SNR imbalance, although affecting the system performance,
does not change the diversity order. We then show by simulation that
channel estimation error has a large effect on the cooperative relay
assisted OSTBC communication system and deserves more attention when
designing the system.

\section*{Appendix}
\subsection{Derivation of \eqref{eq:Pe}}
The error probability is given by
\begin{eqnarray}
Pe=E\left(Q\left(a\sqrt{\Upsilon}\right)\right)
\label{eq:Pe_general}
\end{eqnarray}
where $\Upsilon=\frac{1}{1+r}\gamma\left(|h_B|^2+r|h_R|^2\right)$ is
the instantaneous received SNR and a summation of two independent
central Chi-square distributions; $Q(\cdot)$ is Gaussian Q function
defined as
\begin{eqnarray}
Q(x)=\frac{1}{\pi}\int_0^\frac{\pi}{2}\exp\left(-\frac{x^2}{2\sin^2\theta}d\theta\right),\quad
x\geq0.
\end{eqnarray}
Thus we have \eqref{eq:Pe_general} as
\begin{eqnarray}
Pe\!=\!\!E\!\!\left(\frac{1}{\pi}\int_0^\frac{\pi}{2}\!\!\exp\!\!\left(\!-\frac{a^2\frac{1}{2}\gamma\left(\frac{2}{1+r}|h_B|^2\!+\!\frac{2r}{1+r}|h_R|^2\right)}{2\sin^2\theta}d\theta\!\!\right)\!\right).
\end{eqnarray}
By changing the order of averaging and integration ( Eq
(3.54),\cite{Tse:book_05}) and averaging with respect to $h_B$ and
$h_R$ under the independent Rayleigh fading assumption, we get
\begin{eqnarray}
Pe\!\!\!\!&=&\!\!\!\!\frac{1}{\pi}\!\int_0^\frac{\pi}{2}\!\!\left(\frac{1}{1+\frac{a^2\frac{1}{2}\gamma\frac{2}{1+r}}{2\sin^2\theta}}\right)\!\!\!\!\left(\frac{1}{1+\frac{a^2\frac{1}{2}\gamma\frac{2r}{1+r}}{2\sin^2\theta}}\right)\!\!d\theta\nonumber\\
\!\!\!\!&=&\!\!\!\!\frac{1}{\pi}\frac{1}{
MN}\frac{1}{\frac{1}{N}\!-\!\!\frac{1}{M}}\!\!\int_0^\frac{\pi}{2}\!\!\left(\frac{1}{\frac{1}{\sin^2\theta}\!\!+\!\!\frac{1}{M}}\!-\!\!\frac{1}{\frac{1}{\sin^2\theta}\!\!+\!\!\frac{1}{N}}\right)d\theta,\label{eq:mid}
\end{eqnarray}
where $M=\frac{a^2\gamma}{2(1+r)}$ and $N=\frac{a^2
r\gamma}{2(1+r)}$.

It is pointed out in \cite{Lin:ISITA_04}\cite{Verdu:book} and
\cite{Knopp:IT_00} that
\begin{eqnarray}
\frac{1}{\pi}\int_0^{\frac{\pi}{2}}\frac{1}{\frac{1}{\sin^2\theta}+\frac{1}{M}}d\theta=\frac{1}{2}M\left(1-\frac{1}{\sqrt{1+\frac{1}{M}}}\right),
\end{eqnarray}
thus \eqref{eq:mid} becomes
\begin{eqnarray}
Pe&=&\frac{1}{2MN\left(\frac{1}{N}-\frac{1}{M}\right)}\left(M\left(1-\frac{1}{\sqrt{1+\frac{1}{M}}}\right)\right.\nonumber\\
&&\left.-N\left(1-\frac{1}{\sqrt{1+\frac{1}{N}}}\right)\right)\label{eq:mid2}
\end{eqnarray}
Assume $\mu_M=\sqrt{1+\frac{1}{M}}$ and
$\mu_N=\sqrt{1+\frac{1}{N}}$, we have $\frac{1}{M}=\mu_M^2-1$ and
$\frac{1}{N}=\mu_N^2-1$; $\mu_M=\sqrt{1+\frac{2(1+r)}{a^2\gamma}}$
and $\mu_N=\sqrt{1+\frac{2(1+r)}{a^2\gamma r}}$,

The probability of error in \eqref{eq:mid2} is
\begin{eqnarray}
Pe&=&\frac{(\mu_M^2-1)(\mu_N^2-1)}{2(\mu_N^2-\mu_M^2)}\left(\frac{1}{\mu_M^2-1}\left(1-\frac{1}{\mu_M}\right)\right.\nonumber\\
&&\left.-\frac{1}{\mu_N^2-1}\left(1-\frac{1}{\mu_N}\right)\right)\nonumber\\
&=&\frac{1}{2}(1-\frac{1}{\mu_M})(1-\frac{1}{\mu_N})(1+\frac{1}{\mu_M+\mu_N}).\label{eq:mid3}
\end{eqnarray}

Plugging $\mu_M=\sqrt{1+\frac{2(1+r)}{a^2\gamma}}$ and
$\mu_N=\sqrt{1+\frac{2(1+r)}{a^2\gamma r}}$ into \eqref{eq:mid3}
completes the proof \hfill$\blacksquare$

\bibliographystyle{IEEEtran}
\bibliography{SNR_imb}
\end{document}